\documentclass[prb, showpacs, preprint]{revtex4}
\usepackage{graphicx,epsfig,rotate}
\usepackage{dcolumn}%
\usepackage{bm}

\begin{document}


\title{Mechanism for BCC to HCP Transformation:
Generalization of the Burgers Model}

\author{S.G. Srinivasan}
\altaffiliation[Present Address: ]{Materials Science Division, Los
Alamos National Laboratory,
Los Alamos, New Mexico 87545 (sgsrini@lanl.gov).}

\author{D.M. Hatch}
\altaffiliation[Permanent Address: ]{Department of Physics and Astronomy,
Brigham Young University, Provo, Utah 84602 (hatchd@byu.edu).}

\author{H.T. Stokes}
\altaffiliation[Permanent Address: ]{Department of Physics and
Astronomy, Brigham Young
University, Provo, Utah 84602.}

\author{A. Saxena}
\author{R.C. Albers}
\author{T. Lookman}
\affiliation{Theoretical Division, Los Alamos National Laboratory, Los
Alamos, New Mexico 87545.}

\date{\today}


\begin{abstract}
Many structural transformations involve a group-nonsubgroup
relationship between the
initial and transformed phases, and hence are beyond the purview of
conventional
Landau theory. We utilize a systematic and robust methodology to describe such
reconstructive martensitic transformations by coupling
group-theoretical arguments to
first-principles calculations. In this context we (i) use a
symmetry-based algorithm to
enumerate transformation paths, (ii) evaluate the energy barriers along these
transformation paths using all-electron first principles
calculations, (iii) deduce
the full set of primary and secondary order parameters for each path
to establish the
appropriate Ginzburg-Landau free-energy functionals, and (iv) for each
path, identify
special points of the primary order parameter, as a function of
local distortions,
corresponding to the end product phase. We apply this method to the study of a
pressure driven body-centered cubic (bcc) to hexagonal close-packed (hcp)
transformation in titanium. We find a generalization of the Burgers
mechanism, and
also find that there is no energy barrier to this transformation.
In fact, surprisingly,
we also find a region of volumes in which the intermediate path
becomes more stable
than either of the end-points (bcc or hcp).  We therefore predict a
new orthorhombic phase
for Ti between 51 and 62 GPa.

\end{abstract}

\pacs{81.30.Kf, 64.70.Kb, 05.70.Fh}

\maketitle


\section{Introduction}
Polymorphism, {\it i.e.}, the changing of a solid from one crystal
structure variant
to another, is abundant in both nature and in engineering technologies. The
body-centered-cubic (bcc) to hexagonal-close-packed (hcp) structural
transition
discussed in this paper is a well-known example. For instance, iron
undergoes this
transformation at very high pressure ($\sim13$ GPa) in the core of
the earth as well
as under shock conditions in diamond anvil experiments \cite{iron}.
Certain seismic
activity has been attributed to strain generated during this phase
change. Toughness
and ductility in Ti and Zr, which undergo this transformation at
high temperature,
are two important issues for the aerospace industry. In addition to
these three elements,
the bcc-hcp transformation is found in eighteen other elements, such
as Ba, Mg, Hf, Sc, Tl,
and Y, and their alloys.

Polymorphic diffusionless transformations have been broadly
classified as either
displacive or reconstructive. In the displacive transformation there
is no change in
the first coordination of the atoms in both the initial and the
transformed phases.
The energy differences between these two phases arise from changes
in the secondary
coordination. In terms of crystal symmetry, the transformed phase is
a slightly
altered derivative crystal structure, a subgroup, of the initial phase. In the
reconstructive transformation, bonds are broken and then reformed to suit the
transformed phase. Consequently, there is a change in the primary
coordination. The
transformed phase is no longer a slightly altered structure of the
initial phase; it is
not a subgroup structure of the parent phase (or, vice-versa). In
diffusionless
polymorphic transformations, individual atoms execute ordered and
correlated motions,
and significant transformational strains may  also develop in this
process. These are
the so-called martensitic transformations (MT).

The study of MT has a long and rich history. Early attempts by metallurgists to
understand both displacive and reconstructive MT's ignored the need to maintain
coherency between the initial and transformed structures as the transformation
progressed. That neglect hindered the development of an atomistic understanding of the
transformation mechanism in terms of simple shears of the crystal cell coupled with
local atom readjustments (shuffles). Thus, martensite theories often chose the
phenomenological route of geometrically relating  the initial and transformed phases,
ignoring the atomistic pathways and microscopic mechanisms driving the transformation.
A well-known phenomenological theory of Wechsler, Lieberman and Read successfully
demonstrated that a MT can be understood at this level in terms of three basic
deformations \cite {christian,Reed-Hill,Wayman}. Nevertheless, within the above
mentioned phenomenological approach, microscopic mechanisms were identified for many
reconstructive transformations, purely from the orientational relationships between
the initial and transformed phases. In Table \ref{reconstMech} we give examples of
several common mechanisms.

More recently, physicists have successfully used the phenomenological Landau theory to
model displacive MT for a nearly continuous ($i.e.$, weakly first order)
transformation. Because the displacive MT is nearly continuous, the invariant
free-energy (invariant under the symmetry group of the initial phase) defines an order
parameter (OP) onset. The Landau philosophy has recently been extended to a
phenomenological theory of reconstructive phase transitions by expressing the OP
function as a density-wave function of the atomic displacements \cite{Toledano}.  Note
that an effective {\it one-dimensional} Ginzburg-Landau model for Ti and Zr
\cite{lindgard} has been combined previously with electronic structure calculations
\cite{sanati}. The bcc-hcp transition in titanium occurs around 1150 K. However, here
we attempt to model this transition at $T=0$ by varying the pressure: bcc is the
high-pressure phase and the hcp the low-pressure phase.   The bcc-hcp transformation
pathways and mechanisms elaborated below are generic to the bcc-hcp transition in a
wide variety of elements and alloys.  In effect, the temperature variation in titanium
is mimicked by pressure variation at $T=0$ capturing the essential pathways of the
transition, specifically the intermediate orthorhombic phase. As discussed below, we
surprisingly find that one of the pathways becomes more stable than either the bcc or
hcp end-points in Ti for a range of volumes.  We therefore predict a new orthorhombic
phase transformation in Ti at low temperature.  This paper is organized as follows. In
the next section we introduce the concept of a common subgroup.  In Sec. III we
enumerate the various symmetry based pathways for the bcc-hcp phase transition. In Sec.
IV we describe the first-principles calculations performed here. Section V gives an
overview of the Landau free-energy fitting. Results and discussions are contained in
Sec. VI.

\section{The common subgroup paradigm and transformation paths}
The bcc-hcp transformation in Ti considered here is a reconstructive martensitic
transformation. There is no group-subgroup relationship between the space groups of
the bcc ($Im\bar3m$) and hcp ($P6_3/mmc$) phases participating in the transformation.
Burgers \cite{burgers} originally proposed the orientational relationship between the
bcc and hcp phases in Zr: the basal $\{{0001}\}_{\rm hcp}$ plane is parallel to a
$\{{110}\}_{\rm bcc}$ plane, and a $\langle 11\bar{2}0\rangle$ close-packed direction
in the hcp basal plane is parallel to a cubic $\langle 111\rangle$ direction. A
combination of three lattice distortions can lead to this transformation
\cite{petry,heiming,tramp}: (i) a shuffle displacement of the $\{{110}\}_{\rm bcc}$
planes in the $\langle 1\bar{1}0\rangle$ directions. A displacement amplitude of
$\sqrt{2} \over 12$ times the bcc lattice constant leads to the exact hcp stacking
sequence. (ii) A shear such as (1$\bar{1}$2) [$\bar{1}$11] and
($\bar{1}$12)[1$\bar{1}$1] to squeeze the bcc octahedron into a regular hcp one. The
angle of the hcp face in the basal plane changes  from 70.53$^\circ$ to 60$^\circ$.
The Burgers mechanism was painstakingly determined by studying the orientational
relationships between bcc and hcp phases. It is understood in terms of the three
parameters, shuffle, shear, and volume dilatation.

In this paper we present a more general and robust approach for identifying
reconstructive transition mechanisms based on crystallographic group theory. It is
equally applicable to bcc-hcp MT's in other elements, or for other reconstructive MT's
such as the transitions fcc-hcp, hcp-$\omega$, etc. The distortion of a crystal can
clearly destroy some of its symmetry elements and force it into one of its lower
symmetry subgroups. As atoms move from one structure toward another, symmetry is
lowered as the atoms leave their high symmetry locations.  We assume initially that
this lower symmetry is maintained until at some suitable, large displacement it is
able to lock-in to the symmetry of the transformed phase. We thus hypothesize that the
reconstructive transformations proceed via a subgroup common to both the initial and
transformed phases. This requires the atoms to sensibly map from the initial structure
onto the subgroup, and then onto the final structure.  By imposing some restrictions
on cell size changes the number of available transformational paths equals a finite
number of maximal subgroups \cite{SH-Tables}, with compatible atomic mappings, common
to both the initial and the transformed structures. We can then look for even lower
symmetry structures along the path by testing stability within the maximal subgroup
phase. The actual path chosen is ultimately decided by the nature of the interatomic
interactions in the material. Coupling the symmetry with interatomic force
calculations gives us a general, systematic, and robust recipe to explore the
different pathways, and identify the minimum energy path(s) among them. This method is
amenable to automation by computer and can quickly search for transformational paths
between any two of the 230 crystallographic space-groups.

We illustrate an application of our procedure by coupling  the full potential linear
augmented-plane-wave (FLAPW) method to crystal symmetry information in order to
enumerate all possible paths for the bcc-hcp transformation in titanium. Based just on
symmetry conditions, we initially obtain six possible mechanisms for the bcc-hcp
transformation in titanium. Of these six, one is energetically favored and is similar
to the Burgers mechanism. Our mechanism adds an effective interplanar variance to the
three parameters used in the original Burgers mechanism. In this work, we extend the
Landau philosophy to reconstructive MT's by including local site symmetry
information.  Our procedure is distinct from that given in Toledano et al.
\cite{Toledano} in that we do not assume a transcendental stationary density-wave
description but instead structure our approach on common subgroup symmetry aspects and
a more conventional Landau position-dependent OP. However, we do use specific
intracell correspondences of atoms in the two structures, which is not contained in
the conventional Landau procedure, and look for the onset of new symmetry elements at
the end product phase.  These features are in common with the description given by
Toledano et al. \cite{Toledano} We fit a Landau free-energy functional over the
multi-component OPs in this transition. The free-energy functional is fitted using the
first principles energy values obtained for the most plausible transformation
mechanism.

The bcc-hcp transition in titanium occurs around 1150 K. Here we attempt to model this
transition at $T=0$ by varying the pressure: bcc is the high-pressure phase and hcp
the low-pressure phase.  In this sense our modeling cannot be considered as realistic
in capturing the (temperature induced) transition in titanium. However, two new
high-pressure phases of titanium have been observed recently, namely $\gamma$-Ti which
is a distorted hcp structure at 128 GPa \cite{vohra,akahama} and $\delta$-Ti which is
a distorted bcc structure at 140 GPa \cite{akahama}.  Interestingly, both of these
structures are orthorhombic with the space group $Cmcm$ and the lattice deformations
can be interpreted within the framework of our generalized Burgers mechanism. We also
predict a new orthorhombic phase transformation between 51 and 62 GPa at zero
temperature. In this context our modeling is directly relevant to pressure induced
transitions in titanium. In addition, the bcc-hcp transformation pathways and
mechanisms elaborated below are representative of the bcc-hcp transition in a wide
variety of elements (e.g. Zr, Hf) and alloys. In effect, the temperature variation in
titanium is mimicked by pressure variation at $T=0$ capturing the essential pathways
of the transition, specifically the intermediate orthorhombic phase.

\section{Crystal Symmetry Based Algorithm to determine the mechanism
for the bcc-hcp transformation}

Methods based solely on orientational relationships between the initial and
transformed phases force one to ``guess" the transformational mechanism. Even symmetry
based information coupled with atomic site correspondence does not unequivocally
reveal the transformational mechanism. Our approach here is to use ``physics'' insight
(interatomic forces based on first-principles or some other suitable semi-empirical
method) to interrogate the various common subgroup based symmetry pathways to
determine plausible paths based on their relative energies. We expect that the minimum
energy path, under the transformational conditions, will be favored, and we initially
consider maximal symmetry pathways. Overall, such an approach reduces the systematic
search of paths to a smaller number. In this process, we can also identify the primary
and secondary order-parameters, the Ginzburg-Landau invariants, and declare the
special OP lock-in values corresponding to the transformed phase.

To describe a reconstructive phase transition, one must address two important
questions: (1) How are the atoms mapped from one structure onto the other?  (2) What
distortional pathway do the atoms take between these two structures? The mapping
question deals with how the two structures are related. The path question is more
difficult. It deals with actual atomic displacements and strains that occur during the
phase transition. In an earlier paper we presented a systematic procedure for
obtaining possible mappings (paths) for a reconstructive phase transition in two
dimensions (2D) \cite{Hatch}, namely the square lattice to triangular lattice
transition, based essentially on symmetry considerations. In that treatment, it was
assumed the pathway between the two structures proceeded by means of an intermediate
common unstable structure (either an oblique lattice or a centered rectangular
lattice) with definite space-group symmetry $G$. Note that the square$\to$triangle
transformation is a 2D analog of the bcc-hcp transformation.  If we are considering a
reconstructive phase transition from a structure with space-group symmetry $G_1$ to a
structure with space-group symmetry $G_2$, then our description is actually a two-step
process, $G_1\to G\to G_2$, where $G$ is a subgroup of $G_1$ and also a subgroup of
$G_2$ so that each step, $G_1\to G$ and $G\to G_2$, is a transition with a
group-subgroup relationship.

The first question stated above is related to the mapping of atoms
through a subgroup
and was systematically addressed by a four step algorithm
\cite{Hatch}. We describe
this algorithm using the bcc-hcp transformation as an example.
However, it is general
and applies to other reconstructive transformations. The tables of
Stokes and Hatch
\cite{SH-Tables} were used to obtain the needed information at each
of the following
four steps.

$Step$-1: Find the subgroups common
to both the initial (bcc) and transformed (hcp) phases. To begin, it
is convenient to
look at common subgroups arising from distortions corresponding to
${\bf k}$ points of
symmetry. We find seventeen subgroups which are subgroups of both the hcp
and bcc phase.

$Step$-2: Subgroups arise from microscopic displacements and/or macroscopic strains.
Strain cannot be the primary OP since the bcc-hcp transition involves primitive cell
doubling. Shuffle is our primary OP and the induced representation formalism is used
to determine which irreducible-representations (IR's) allow microscopic atomic $x,y,z$
displacements from each parent group into the subgroups. Only fifteen subgroups are
consistent with this requirement.

$Step$-3: Atoms are at the 2(a) and 2(c) Wyckoff positions in bcc and hcp
respectively. Subgroups that relate bcc and hcp phases must have
compatible Wyckoff
positions. Each acceptable subgroup is connected to a specific IR
and mode form (OP
direction). Thus, there is a precise atom to atom identification
throughout the
transformational path. This ensures consistency as the Wyckoff
position changes (or
splits) as the transformation moves from bcc to hcp. After applying
this constraint,
we are left with fourteen subgroups corresponding to fifty-eight
possible mechanisms.

\begin{figure}
\begin{center}
\includegraphics *[width=8.0cm,angle=0]{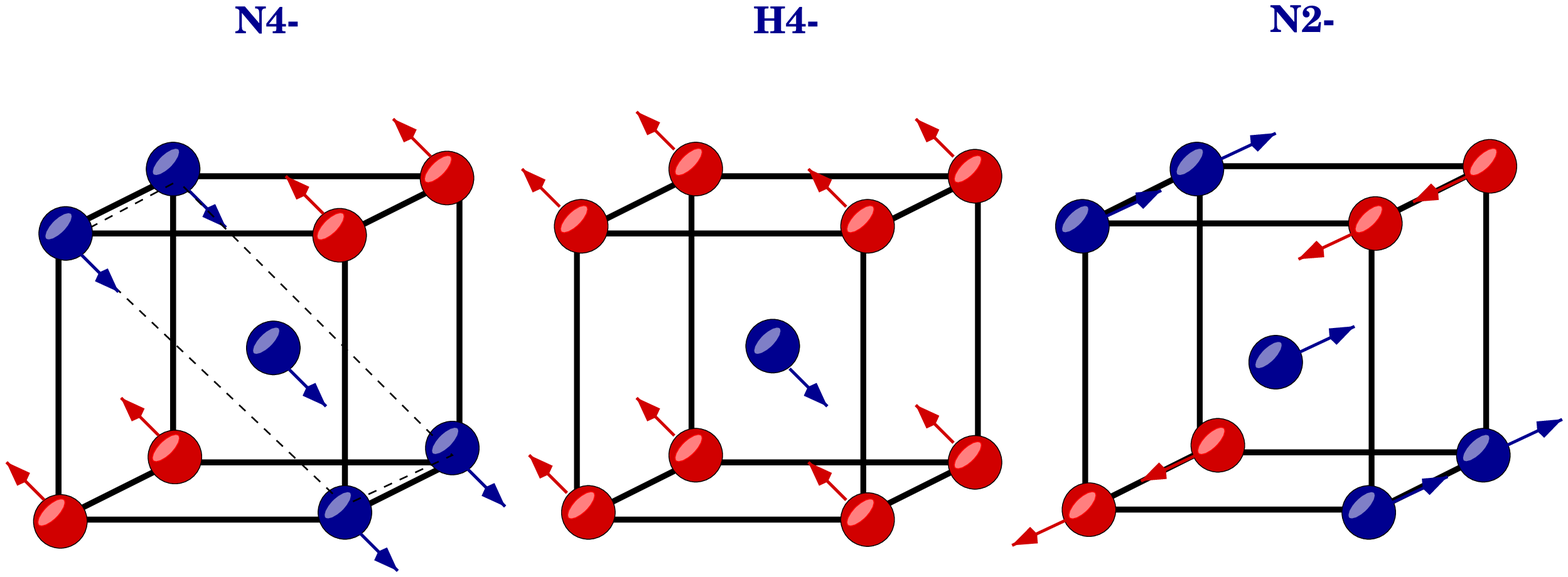}
\caption{Three simplest mechanisms that take a bcc phase into a hcp
structure, and
based on a common orthorhombic subgroup (Space-Group 63, $Cmcm$):
(a) N$_4^-$, (b)
H$_4^-$, and (c) N$_2^-$.} \label{fig:mechBCC_HCP}
\end{center}
\end{figure}

$Step$-4: We determine a specific (correlated) displacement mechanism for each IR and
subgroup. All possible mechanisms for the bcc-hcp transformation based on this
algorithm are given in Table \ref{allMech}.  Of these, we restrict our attention to
the most simple mechanisms such as a maximum cell size change of two and a P1 or P2
(one parameter) OP direction. As can be seen from Table II, these two restrictions
yield three mechanisms; $H_4^-$ 63(P2,2), $N_2^-$ 63 (P1,2), and $N_4^-$ 63(P1,2).
This notation gives the IR, the identifying number of the subgroup, the order
parameter direction, and the cell size change. The atomic displacements for these
mechanisms are shown in Fig. \ref{fig:mechBCC_HCP}.  Note that the space group
symmetry of the common subgroup is the same and the atoms go into the Wyckoff $c$
postions for each of the 3 paths.

The labeling of representations is that of Miller and Love \cite{ML}, the space group
number is according to the International Tables\cite{International}, and the OP
direction is that of Stokes and Hatch\cite{SH-Tables}. Specifically, N and H refer to
the high symmetry points in the bcc Brillouin zone, namely the face center and corner,
respectively. Note that a phonon anomaly has been observed at the N-point in the
phonon dispersion curves of Ti, Zr and Hf \cite{petry,heiming,tramp}. It is
conceivable that the H-point mechanism may be related to the small anomaly seen at the
H-point phonon dispersion of Mg \cite{mg}. The crux of the problem is to couple the
pathway information to the relevant physics in order to identify which are the most
energetically favorable paths.

We generalized the above algorithmic procedure by allowing multiple
primary order
parameter mechanisms and considering additional user input for some important
parameters such as allowed strain, nearest-neighbor distances, and
unit-cell size
change. This computer program is called COMSUBS and filters possible
subgroups through
user defined conditions. The program was first used in the
description of the rock-salt to
CsCl structural change in sodium chloride \cite{Stokes}. We applied
the COMSUBS
algorithm to the description of the transition of interest here, the
hcp to bcc
transition in titanium.

For the low-temperature hcp structure, $G_1=P6_3/mmc$ with lattice parameter
$a_1=2.645$~\AA, $c_1=4.11~\AA$. For the high-temperature bcc structure, $G_2=Im\bar3m$
with lattice parameter $a_2=2.91 ~\AA$. We used the following criteria:

(1) We considered only those subgroups where the length of the lattice
generators is
5.03~\AA\ or less.  This condition effectively limits the size of
the unit cell in the
common subgroup, which in turn limits the size of the allowed unit
cell parameters.

(2) We considered only those subgroups where the principal elements of the
strain tensor are
less than $1+\epsilon$ and greater than $(1+\epsilon)^{-1}$, where
$\epsilon=0.5$.
This condition limits the strain (lattice vector lengths) to be
within this allowance
and also the angle relationships between the group and its subgroup.

(3) The nearest-neighbor distance is 2.56~\AA\ in $G_1$ and
2.52~\AA\ in $G_2$.  We
considered only subgroups where the nearest-neighbor distance in the
structure halfway
between $G_1$ and $G_2$ is greater than 2.03~\AA\ (80\%\ of the
average of 2.56~\AA\
and 2.52~\AA).  If the nearest neighbor distance is less than our
chosen distance of
2.03~\AA\ we reject the path.  Along a rejected path the lattice
would need to expand
to allow atoms to pass one another and we assume such a strain
implies an energy
barrier unfavorable to this path.

(4) We considered only maximal subgroups.  These define the possible mappings of atoms
in $G_1$ onto atoms in $G_2$.  Subgroups of these maximal subgroups do not introduce
new mappings.  They only alter the path by allowing additional distortions in $G$
along the path.  We check for these additonal distortions when a maximal subgroup is
selected.

The criteria given above were chosen somewhat subjectively.  Each of the criteria could
be relaxed and the list would thus be extended.  We will see below that the values we
chose resulted a list which contained a subgroup that upon initial screening and
additional checks for stability is the likely candidate for the pathway mechanism.

 Using these criteria, we obtained six subgroups from COMSUBS, which we list in Table
\ref{en6Mech}. For the first two entries in the table there is no change in the size
of the primitive unit cell relative to the hcp cell. The next two entries are
subgroups where the size of the primitive unit cell is doubled.  The last two entries
are subgroups where the size of the primitive unit cell is three times larger than
that of the hcp cell. Subgroups with larger primitive unit cells were not found since
we limited the length of the generators according to criterion (1) above. We label the
six potential mechanisms we obtained as, (a) ort63A, (b)ort63B, two pathways via
orthorhombic space-group 63, (c) mon14, a pathway via the monoclinic space-group 14,
(d) mon15, via the monoclinic space-group 15, (e) mon9, via the monoclinic space-group
9, and (f) tri2, via the triclinic space-group 2.

   The list obtained by COMSUBS does contain
groups in common with those obtained from the four-step algorithm described earlier
(for example ort63B), some are excluded by COMSUBS since they do not meet the criteria
listed above, and new pathways are obtained due to the allowance of coupled parameters
(for example mon14 couples $N_4^-$ and $N_2^-$).

Prescreening calculations showed (see Table \ref{en6Mech}) that the second pathway
based on the orthorhombic space-group Cmcm(63) had the lowest energy barrier among all
these paths, and it was chosen for detailed study. The calculations were done at
midpoint structures as determined by the bcc-hcp endpoint structures. This
distortional mechanism is described in terms of an appropriate high-symmetry point (N)
of the bcc Brillouin zone and irreducible representation N$_4^-$, with alternating
shuffles in the $\pm$[0$\bar{1}$1] directions in the (011) planes. This mechanism is
closely related to the original Burgers mechanism.

\section{Mechanisms for the bcc-hcp transformation in Titanium}
The most favored transformation path should have the lowest energy
among all the paths considered here. Energy of the different paths
were computed using an all-electron, full potential linear
augmented-plane-wave (FLAPW) method \cite{wien97}. Calculations were
performed scalar-relativistically, neglecting spin-orbit coupling for
the valence electrons. Local orbitals \cite{singh} were added to
enhance the variational freedom and allow the semi-core $3s$ and $3p$
orbitals to be treated along with the valence electrons. The added
energy parameter was used to simultaneously treat the residual $s$
and $p$ character of the valence electrons. Our results are
insensitive to small changes in all energy parameters.

Our calculations used the generalized gradient approximation (GGA)
of Perdew, Burke,
and Ernzerhof \cite{pbe} for the exchange-correlation functional. A
muffin-tin sphere
with a radius $R_{\rm MT}$ was used to define the augmented
plane-wave basis functions.
No restriction is made as to the shape of the potential or charge
density. The size of
the FLAPW basis was determined by a plane-wave cutoff, $K_{max}$.
The $K_{max}$ value
is given by the relation $R_{MT} K_{max} = 9.0$, and was found to be
adequate in
determining the system energy. $405$ irreducible points in the first
Brillouin zone of
orthorhombic Ti were sampled according to the improved tetrahedron method
\cite{blochl} for Brillouin zone (BZ) integrations. Total energies
changed by less
than $0.014$ eV/atom on increasing the number of BZ points sampled. All
self-consistent calculations were iterated until the total energy
changed by less than
$0.1$ meV/atom.

\begin{figure}
\begin{center}
\includegraphics*[width=8.0cm,angle=270]{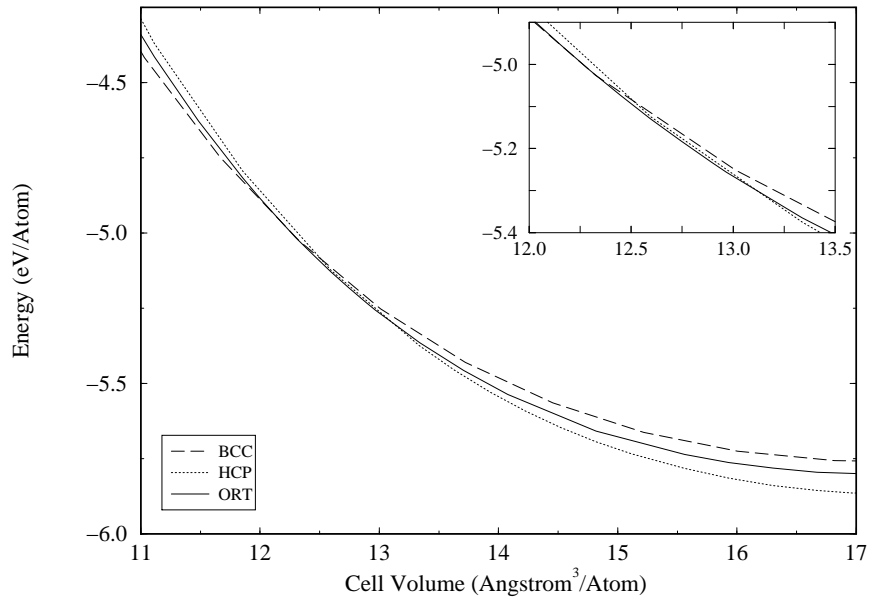}
\caption{Energy versus cell volume for the bcc, hcp and
orthorhombic phases of titanium. The orthorhombic phase has the lowest
energy within the transformation volume region.}
\label{fig:enBCC_HCP}
\end{center}
\end{figure}

Based on the FLAPW calculations, we found that an intermediate orthorhombic pathway
becomes more stable than either of the bcc or hcp end-points for volumes between
49.197 and 50.235 ${\rm \AA}^3$. The intermediate orthorhombic phase gives rise to
orthorhombic-bcc and hcp-orthorhombic transitions at 61.8 and 51.3 GPa, respectively
(See
Fig. \ref{fig:enBCC_HCP}). Because the orthorhombic unit cell and the atoms
within it were relaxed to find the lowest energy, these parameters change slightly in
this range.  In Table \ref{ortParams}, we list the relevant atomic and unit cell
coordinates near the middle of the stable orthorhombic region. The $y$ coordinate
refers to the value of $y$ in the specification of the Wyckoff $c$ position of the
form $0,y,1/4$. If we ignore this orthorhombic intermediate phase, we find that the
bcc-hcp transition occurs at a constant pressure of $\sim$ 57 GPa. The transition
pressures were determined by the slope of the common tangent to the energy versus
volume curves of the two appropriate phases participating in the transformation. Since
the curves are very close to each other in energy, we are unable to determine from the
common tangent construction whether the orthorhombic phase is more stable than a
mixture of bcc and hcp crystals. We do see that there are regions wherein the
orthorhombic system has lower energy than either the hcp or bcc phases.

Since our preliminary calculations pointed to a mechanism operating via a common
orthorhombic subgroup, we chose to investigate it in more detail. The orthorhombic
cells were fully relaxed with respect to its lattice parameters and internal
coordinates, a total of four variational parameters, to determine minimum energy
configurations. Such minimum energy surfaces for the common orthorhombic subgroup were
determined at five different volumes, 49.316, 49.434, 49.716, 50.054, and 50.235 ${\rm
\AA}^3$, in the transformation region. We find that the energy of the orthorhombic
phase is also sensitive to its lattice constant along the $c$-axis. All of the five
volumes possessed minimum energy when the orthorhombic cell had a length between 4.04
and 4.07 $\AA$. This $c$-axis value is significantly different from the value of 4.11
$\AA$ corresponding to the clamped orthorhombic structure of the hcp phase. Such a
variation in the $c$-axis lattice constant was not considered in the original Burgers
mechanism. The energy versus volume plot for the bcc, hcp and orthorhombic phases in
the vicinity of the transformation region is given in Fig. \ref{fig:enBCC_HCP}. We can
see that there is no effective energy barrier to the bcc-hcp transformation when the
transformation proceeds via the orthorhombic phase.


\section{Determining Free-Energy Functional From
First-Principles Data}

Here we demonstrate that it is possible to rigorously fit the Ginzburg-Landau free-energy functional to first-principles energy data.  These functionals will be used in our future work. One primary and three secondary order parameters (OP's)
completely describe the bcc-hcp transition through the orthorhombic phase. They are the
intracell atom ``shuffle'' ($\eta_1$), the shear ($\eta_2$), deviatoric ($\eta_3$),
and volumetric ($\eta_4$) strains. Shuffle is the primary OP and is in units of Bohr.
The orthorhombic $c$-axis parameter gives us a measure of $\eta_2$, and is also in the
units of Bohr. The deviatoric OP $\eta_3$, defined as the ratio of $a$- to $b$-axis
parameters of the orthorhombic cell, is dimensionless. The volumetric OP ($\eta_4$)
has the units of Bohr$^3$. The full form of the Ginzburg-Landau free-energy
functional, involving all four OP's, is given by the equations:
\begin{eqnarray}
\label{eq:glE}
F_L(\eta_1,\eta_2,\eta_3,\eta_4) = A_1{\eta_1}^2 + A_2{\eta_1}^4 +
A_3{\eta_1}^6  \nonumber\\
+ A_4{\eta_1}^2{\eta_2}^2 + A_5{\eta_2}^2 + A_6{\eta_2}^4  \nonumber\\
+ A_7{\eta_1}^2{\eta_3}^2 + A_8{\eta_1}^2{\eta_2}
+ A_9{\eta_3}^2 \nonumber\\
+ A_{10}{\eta_3}^4 + A_{11}{\eta_1}^2{\eta_3}
+ A_{12}{\eta_4}^2 \nonumber\\
+ A_{13}{\eta_4}^3 + A_{14}{\eta_2}^2{\eta_3}
+ A_{15}{\eta_3}^3 \nonumber\\
+ A_{16}{\eta_1}^2{\eta_4} + A_{17}{\eta_2}^2{\eta_4} +
A_{18}{\eta_3}^2{\eta_4},
\end{eqnarray}

\begin{eqnarray}
\label{eq:gE}
  F_G(\eta_1) = g_1{\eta_{1,x}}^2 + g_2({\eta_{1,y}}^2 +
{\eta_{1,z}}^2) 
+ g_3{\eta_{1,y}}{\eta_{1,z}}.
\end{eqnarray}
\noindent Equation (\ref{eq:glE}) is the Landau portion of the free
energy and gives
the energy for a
homogeneous phase arising from coupled primary and secondary OP
contributions. Equation (\ref{eq:gE})
contains the (nonlocal gradient) Ginzburg portion of the free energy
and gives the
energy of heterogeneous solutions, e.g., energies of domain walls
configurations,
etc.  The notation in Equation (\ref{eq:gE})  indicates derivatives
of the OP components,
e.g., $\eta_{1,x}$ is the $x$ derivative of $\eta_{1}$.

It is important to note that this free-energy is at $T=0~\rm K$.  The coefficients of
$F_L$, the Landau portion of the free energy given above, were determined by a Powell
non-linear least square fitting \cite{recipe} of the first-principles energy data,
along the bcc-hcp transition path. Since the purpose of this work is to demonstrate
the feasibility of fitting GL free-energy functional to first-principles energy data,
we carried out the fit at one fixed volume (50.054 \AA$^3$) falling within the
transformational volume region and using 58 energy values corresponding to this volume
obtained from first-principles calculation. The coefficients are given in Table
\ref{glCoeff}. The fitted energy values had an error of less than 0.5 $\%$. Here, the
summation index $i$ is over all the  58 energy values.

\section{RESULTS, DISCUSSION and CONCLUSION}
As introduced earlier, Burgers \cite{burgers} originally derived the
orientational
relationship between the bcc and hcp phases for the bcc-hcp
transformation. This
orientational relationship was translated into a mechanism
\cite{petry,heiming,tramp}
with three parameters: (a) shuffle of the atoms in the
$\{{110}\}_{\rm bcc}$ planes in
the $\langle 1\bar{1}0\rangle$ directions; a displacement amplitude
of $\sqrt{2} \over
12$ times the bcc lattice constant leads to the exact hcp stacking
sequence, (b) a
shear such as (1$\bar{1}$2) [$\bar{1}$11] and
($\bar{1}$12)[1$\bar{1}$1] to squeeze
the bcc octahedron to transform it into a regular hcp one; this
changes the angle of
the hcp face in the basal plane from 70.53$^\circ$ to 60$^\circ$, and (c) an
appropriate volume dilatation to reach the correct end volume. Such
a mechanism is
similar to our pathway via the intermediate orthorhombic subgroup in
the above three
respects. However, our pathway has an additional characteristic in
that it also allows
the orthorhombic lattice constant along the c-axis to vary. In
retrospect, although
such a variation is intuitive, it is surprising that Burgers
apparently did not factor
such a possibility into his mechanism. We suggest that the
material should follow our
more general four parameter pathway during the transformation. In
essence, the utility
of our coupled symmetry and first-principles calculations is
demonstrated in that we
find a Burgers like mechanism, albeit a more generalized one. We
used a general and
robust algorithm to ferret out the correct mechanism, one that is
consistent with the
symmetry and physics of the transformation. There is no energy
barrier for this new
generalized Burgers mechanism.

We have carried out our calculations for $T=0$ which may be more accurate for pressure
induced transitions such as $hcp\rightarrow\omega$ in titanium. However,
$bcc\rightarrow hcp$ is a thermal transformation which takes place around 1150 K.
Therefore, realistically finite temperature phonon effects \cite{rudin} cannot be
overlooked. We, however, calculated the energies of various frozen phonon structures
(using the computer program FROZSL-INIT by L.L. Boyer and H. T. Stokes) that deviate
from our calculated orthorhombic pathway and list the relative energies of these
distorted structures with respect to the reference orthorhombic energy in
Table~\ref{frozenPhonon}. It is clear that the energy of these phonon distorted
structures are higher than that of our orthorhombic pathway. Thus, we expect the
overall transformation path to remain Burgers-like at finite temperature  except that
energy barriers for competing paths may change ($\sim$100 meV for this temperature).
Work is underway to add the effects of entropy in the free-energy. In addition,
defects are invariably present during many phase transformations. Depending on the
defect energetics and concentration they may alter the transformation path or in some
cases arrest it altogether (e.g. a few ppm oxygen arrests the shock-induced
$hcp\rightarrow\omega$ transformation in titanium).  For low energy and low
concentration of defects we do not expect our results to change significantly.
However, there is a need for systematic evaluation of how defects affect transformation
paths and barriers.

To our knowledge this is the first systematic attempt to enumerate
transformation
paths based on symmetry and then to integrate this information with
first principles
calculations for the bcc to hcp transformation.  Our approach is general and
mathematically rigorous with no {\it ad hoc} or empirical elements
involved. We
believe that the unique combination of symmetry and electronic
structure is an equally
powerful tool to identify transformation mechanisms in many other
reconstructive
transformations in nature, e.g., the Bain mechanism for fcc to bcc
transformation
\cite{bain}, the Wentzcovitch-Lam mechanism for fcc to hcp
transformation \cite{wlam},
and the diamond to NaCl structure.  As we have seen for Ti, this
approach also has the
possiblility of predicting new, unforseen phases that would be
difficult to find by
{\it ad-hoc} guesses. Finally, in addition to a Burgers-like
mechanism we predicted
new mechanisms, N$^-_2$ and H$Ø_4$, which may well exist in other
elements and alloys
including Mg.

This research is supported by the Department of Energy under contract
W-7405-ENG-36; it used resources of the National Energy Research
Scientific Computing Center, which is supported by the Office of
Science of the U.S. Department of Energy under Contract
No. DE-AC03-76SF00098.

\newpage

\begin{table}
\caption{\label{reconstMech} Reconstructive transformation mechanisms
from the literature.}
\smallskip
\begin{tabular}{||ll|ll||}
\hline \hline
Transformation &&& Mechanism\\
\hline
bcc--fcc&&&Bain \cite{bain}\\
bcc--hcp&&&Burgers \cite{burgers}\\
bcc--$\omega$&&&Cook \cite{cook}\\
fcc--hcp&&& Wentzcovitch-Lam \cite{wlam}\\
hcp--$\omega$&&&Silcock \cite{silcock}\\
\hline \hline
\end{tabular}
\end{table}

\begin{table}
\caption{\label{allMech} List of all 58 possible mechanisms for the bcc-hcp
transformation allowed by symmetry. For each IR, we specify the common subgroup, OP
direction, and cell size change. The notation for order parameter direction is that of
Stokes and Hatch\cite{SH-Tables}. }
\smallskip
\begin{tabular}{||l|l||}
\hline \hline
IR & Subgroup (OP direction, Cell Size Change)\\
\hline
$H_4^-$ & 11 (C1,2), 63 (P2,2) \\
\hline
$N_2^-$& 1 (S4,4), 1 (6D1,8), 2 (C2,4), 2 (S3,8), 2(4D1, 8) \\
& 5 (S12,8), 5 (4D3,8), 8 (4D6,8), 9 (S13,8), 9(4D4, 8) \\
& 12 (C8,8), 15 (S6,8), 42 (S14,8), 43 (C19,8), \\
& 59 (C1,4), 63 (P1,2)\\
\hline
$N_3^-$ & 1 (S3,4), 1 (6D1,8), 2 (C2,4), 2 (S4,8), 2 (4D1,8) \\
  & 5 (C8,4), 5 (S12,8), 5 (4D3,8), 5 (4D6,8), 8 (S13,8) \\
  & 8 (4D4,8), 12 (C9,8), 12 (S6,8), 42 (C19,8) \\
\hline
$N_4^-$ & 1 (S4,4), 1 (6D1,8), 2 (C2,4), 2 (S3,8), 2 (4D1,8) \\
  & 5 (S12,8), 5 (4D3,8), 8 (S13,8), 8 (4D4,8) 9 (4D6,8) \\
  & 12 (S6,8), 15 (C8,8), 42 (C19,8), 43 (S14,8), \\
  & 58 (C1,4), 63 (P1,2) \\
\hline
$P4$ & 1 (6D,4), 2 (S3,4), 5 (S12,4), 5 (S18,4), 5 (4D1,4) \\
  & 8 (4D6,4), 9 (S5,4), 12 (C8,4), 15 (C2,4), 44 (C22,4) \\
\hline\hline
\end{tabular}
\end{table}

\begin{table}

\caption{\label{en6Mech} Energies from initial pre-screening calculations for six
mechanisms given by the COMSUBS algorithm.  Calculations were done at midpoint
structures as determined by the bcc-hcp endpoint structures. Energy is in eV/atom,
 and the lattice-parameters are in \AA. $^\dag$ denotes an
unrelaxed energy value. All of the rest are relaxed energy values.}

\smallskip
\begin{tabular}{||l|l|l|l||}

\hline \hline
Mehanism & Lattice-Constants & Cell-Angles & Energy \\
\hline
ort63A Cmcm  &   (2.78, 4.35, 4.11)    &  (90, 90, 90) & -4.51 \\
ort63B Cmcm  &   (2.78, 4.35, 4.11)    &  (90, 90, 90) & -5.05 \\
mon14 P21/c  &   (2.57,    4.33, 4.95)    & (90, 116.1, 90)
& -4.84 \\
tri2  P$\bar{1}$     &   (4.77, 4.63, 2.54)    &  (85,
87.2, 62.5) & -4.43 \\
mon9 Cc     &   (10.25, 4.32, 4.47)   &  (90, 48.7, 90) & -4.70$^\dag$ \\
mon15 C2/c  &   (8.18, 4.27, 5.90)  & (90, 46.2, 90)  & -4.24 \\
\hline\hline
\end{tabular}

\end{table}

\begin{table}

\caption{\label{ortParams} Sample atomic and unit cell coordinates for the
intermediate orthorhombic phase. The unit cell parameters are in \AA, the cell angles
are in degrees, and the $y$ coordinate is in fractional atomic units. The $y$
coordinate refers to the value of $y$ in the specification of the Wyckoff $c$ position
of the form $0,y,1/4$. }
\smallskip
\begin{tabular}{||l|l|l|l||}
\hline \hline
Space-Group & Lattice-Constants & Cell-Angles & $y$ coordinate \\
\hline
63 Cmcm  &   (2.92, 4.29, 4.02)    &  (90, 90, 90) & 0.295 \\
\hline\hline
\end{tabular}
\end{table}

\begin{table}
\caption{\label{glCoeff} Ginzburg-Landau free-enegy coefficients in Eq. 1 for the
bcc-hcp transformation in Ti via the ${N_4}^-$ mechanism.  $F_L$ is in eV, $\eta_1$ and
$\eta_2$ are in \AA, $\eta_3$ is dimensionless, and $\eta_4$ is in \AA $^3$.}
\smallskip
\begin{tabular}{||l|l|l||}
\hline\hline
$A_{1{\phantom 0}} = -48106.70$  & $A_{2{\phantom 0}} =  +319.26$ &
$A_{3{\phantom 0}} = +1412.71$  \\
$A_{4{\phantom 0}} = -312.31$  & $A_{5{\phantom 0}} = -848.33$ &
$A_{6{\phantom 0}} = +6.94$ \\
$A_{7{\phantom 0}} = +26707.92$  & $A_{8{\phantom 0}} = +7434.47$ &
$A_{9{\phantom 0}} = -44094.63$\\
$A_{10} = -19221.73$ & $A_{11} = -2145.13$ & $A_{12} = +0.00$ \\
$A_{13} = +0.00$    & $A_{14} = -565.07$ & $A_{15} = +27635.83$ \\
$A_{16} = +0.00$    & $A_{17} = +0.00$  & $A_{18} = +0.00$ \\
\hline\hline
\end{tabular}
\end{table}

\begin{table}

\caption{\label{frozenPhonon} Relative energy ($\Delta$E), in units of meV/atom, of
various frozen-phonon structures that deviate from the transformation pathway via the
common orthorhombic subgroup. The energy of the orthorhombic subgroup is used as
reference. The cell volume is in \AA $^3/\rm atom$. Some IR's have $\it {multiple}$
frozen phonon configurations. Energies of IR's with multiple frozen phonon configurations are listed in the same line. }

\smallskip
\begin{tabular}{||ll|l|ll||l}
\hline \hline
IR &  Space-group &  $\Delta$E \\
\hline
 $\Gamma_1^+$ &    63 $Cmcm$    &       0  \\
 $\Gamma_2^+$ &    11 $P121/m$   &      170 \\
 $\Gamma_3^+$  &    12  $C2/m$    &    1775  \\
 $Y_1^+$     &    51 $Pmma$   &      792 \\
 $Y_2^+$     &    62 $Pnma$   &       9219 \\
 $Y_3^+$     &    58 $Pnnm$   &   71 \\
 $Y_2^-$     &    62 $Pnma$   &    639 \\
 $Y_3^-$     &    57   $Pmmm$ &       391  \\
 $Y_4^-$     &    59 $Pmmn$   &        31 \\
 $Z_1$      &    12  $C2/m$  &       41, 449, 255\\
 $Z_2$      &    15  $C2/c$      &        54 \\
$S_1^+$     &    11 $P121/m$    &     187, 551, 650, 305  \\
$S_1^-$      &    14 $P21/c$  &         332  \\
$S_2^-$     &    11 $P121/m$  &        347, 483, 612 \\
$T_1$      &    12   $C2/m$  &        738, 622, 364, 238  \\
\hline\hline
\end{tabular}

\end{table}




\end{document}